\documentclass[aps,pre,twocolumn,float]{revtex4}

 \usepackage{amsmath,bm,epsfig}

\def \ed {\end{document}}
\def\Fbox#1{\vskip1ex\hbox to 8.5cm{\hfil\fboxsep0.3cm\fbox{%
  \parbox{8.0cm}{#1}}\hfil}\vskip1ex\noindent}  

\newcommand{\eq}[1]{(\ref{#1})}
\newcommand{\Eq}[1]{Eq.~(\ref{#1})}
\newcommand{\Eqs}[1]{Eqs.~(\ref{#1})}
\newcommand{\Ref}[1]{Ref.~\cite{#1}}
\newcommand{\Refs}[1]{Refs.~\cite{#1}}

\def\be{\begin{equation}}\def\ee{\end{equation}}
\def\bea{\begin{eqnarray}}\def\eea{\end{eqnarray}}
\def\bse{\begin{subequations}}\def\ese{\end{subequations}}
\newcommand{\BE}[1]{\begin{equation}\label{#1}}
\newcommand{\BEA}[1]{\begin{eqnarray}\label{#1}}
\newcommand{\BSE}[1]{\begin{subequations}\label{#1}}

\let \nn  \nonumber  \newcommand{\br}{\\ \nn}

\let \= \equiv \let\*\cdot \let\~\widetilde \let\^\widehat \let\-\overline
\let\p\partial

  \def\1{\bm1} 

\def\<{\left\langle}    \def\>{\right\rangle}
\def\({\left(}          \def\){\right)}
 \def \[ {\left [} \def \] {\right ]}

\renewcommand{\a}{\alpha}
\renewcommand{\d}{\delta}
\newcommand{\ve}{\varepsilon}
\renewcommand{\o}{\omega} 
\renewcommand{\L}{\Lambda}\renewcommand{\l}{\lambda}

\def\k{\kappa}

\newcommand{\B}[1]{{\bm{#1}}}
\newcommand{\C}[1]{{\mathcal{#1}}}    

\renewcommand{\sp}[1]{^{\text {#1}}}  
\newcommand{\Sp}[1]{^{^{\text {#1}}}} 
\def\Sb#1{_{\scriptscriptstyle\rm{#1}}}

\begin{document}
\title{Spectrum of  Kelvin-wave turbulence in superfluids}

\author{Victor S. L'vov$^*$  and Sergey Nazarenko$^\dag$}
\affiliation{$^*$ Department
of Chemical Physics, The Weizmann Institute of Science, Rehovot
76100, Israel,\\ $^\dag$ Mathematics Institute, Warwick University, Coventry, CV4 7AL, UK}

\begin{abstract}

We derive a   type of kinetic equation  for Kelvin waves on quantized vortex filaments with random large-scale curvature, that describes step-by-step (local) energy cascade over scales caused  by 4-wave interactions. Resulting new energy spectrum $E\Sb{LN}(k)\propto k^{-5/3}$ must replace in future theory   (e.g. in finding the quantum turbulence decay rate) the previously used spectrum $E\Sb {KS}(k)\propto k^{-7/5}$, which was recently shown to be inconsistent due to nonlocality of the 6-wave energy cascade.
\end{abstract}
\date{\today}
\maketitle

{\bf 1. \emph{Introduction}}. Nowadays, turbulence in superfluids~\cite{1,2} is attracting more and more attention, stimulated by advances in experimental techniques allowing studies of turbulence in various systems
such as $^3$He, ~\cite{3,4}, $^4$He~\cite{5,6} and Bose-Einstein condensates of supercold atoms~\cite{7,8}, and also by an impressive progress in   numerical simulations~\cite{9,10} which give access  to characteristics of turbulence yet unavailable experimentally. One of the most interesting questions is the nature of the  energy dissipation, observed in turbulence of inherently dissipation-free superfluids. This is  especially intriguing  in zero-temperature limit, when normal fluid component disappear together with the obvious dissipation mechanisms: viscosity of normal fluids and mutual friction between the superfluid  and the normal components.

Superfluid turbulence (ST) comprises a tangle of quantized vortex lines which can be characterized by the   intervortex distance $\ell$ and the vortex core radius $a$.   At scales $L\gg \ell$ the discreteness is unimportant and they  can be described classically with the energy flux  toward smaller scales by the    Richardson-Kolmogorov cascade.

Then  the energy is transferred   through the  crossover scale  $\ell$ by some complicated mechanisms~\cite{LNR-1,KS-04}, thereby exciting smaller scales $\ell < \lambda < a$ which propagate along the individual vortex filaments as waves.
These were predicted by Kelvin more than one century ago~\cite{KW}  and experimentally observed  in superfluid $^4$He
about 50 years ago. It is believed  that Kelvin waves (KWs) play a crucial role in superfluid dynamics, transferring energy from $\ell$ to a much smaller scale, where it can dissipate  via emission of bulk phonos.
In a wide range of scales KWs are weakly nonlinear and can be treated within the
theory of weak-wave turbulence~\cite{ZLF}. Such an approach for  KWs was initiated and developed in \cite{KS-04} where a 6-wave kinetic
equation (KE) was presented, and a KW spectrum~\eq{KS} was obtained from this equation based on a dimensional
analysis. This spectrum was subsequently used in theoretical constructions in ST, e.g. to describe the
classical-quantum crossover range of scales and to explain the
ST dissipation rate \cite{LNR-1,KS-04}. However, it was recently shown in \cite{LLNR-09} that spectrum~\eq{KS}
is nonlocal and, therefore, non-realizable. This crucial locality check was only possible after a highly non-trivial
calculation of the 6-wave interaction coefficient done in \Refs{LNR-1}
which took into account previously omitted important contributions.

In this Letter, we exploit the consequences of the nonlocality of the 6-wave theory,  and replace the latter with a new local 4-wave theory
of KW turbulence. Such a 4-wave theory arises from the full 6-wave theory (completed in~\cite{LNR-1}) in the strongly
nonlocal case, when two  waves in the sextet re much longer than the other 4. These waves correspond to the outer scale, - infrared (IR)
cutoff. We  derive a new spectrum of the KW turbulence which is local, and which must be used in future for revising the
parts of the ST where the nonlocal spectrum of the 6-wave theory has previously been used.

{\bf 2. \emph{General background on 4-wave and 6-wave weak turbulence}}.
We begin with a brief overview of weak-wave turbulence for  the 4-wave (of $3 \leftrightarrow  1$ type) and the 6-wave systems, because
these types will be relevant
for our subsequent discussion. Let us start
 with a classical Hamiltonian equation for the complex canonical amplitude of waves $a_{\B k}\equiv a(\B k,t)$ and $a_{\B k}^*$  (classical analogues  of the Bose creations and annihilation operators) with a   wavevector $ \B k$:
\BE{HE}
i  \frac{\p a_{\B k}}{\p t}=\frac{\d \C H}{\d a_{\B k}^*}\ .
\ee
Here $\C H$ is a Hamiltonian,
\be
\label{Hsum}
{\C H} = {\C H}_{\hbox{\small free}} + {\C H}_{\hbox{\small int}}; \;\;\; {\C H}_{\hbox{\small free}} =\int \o_k \, a_{\B k} a_{\B k}^*\,  d {\B k} ,
\ee
where $\o_k$ is the wave frequency. For KWs
  $\o_k={ \L \kappa}\, k^2/{4\pi}$ where $\Lambda =\ln (\ell/a)$ and  $\kappa$ is the circulation quantum.
${\C H}_{\hbox{\small int}}$ is an effective interaction Hamiltonian equal to
 \bea
&& \hskip -.5 cm   \C H_{1\!\leftrightarrow 3}=\!{ \frac 1{6} } \! \int\! \! d \B k_1\dots   d \B k_4 \, \d _{1}^{2,3,4} \big[V_{\,1}^{2,3,4}\,a_1a_2^*a_3^* a_4^* + \mbox{c.c.}\big]\,,  \nn \\ \label{H4}
 && \hskip -.5 cm \C H_{3\leftrightarrow 3}
\!=\! \!
{ \frac 1{36} }\int\! \! d \B k_1\dots   d \B k_6  \d _{1,2,3}^{4,5,6}
  W_{\,1,2,3}^{4,5,6}\,a_1a_2a_3 a_4^*a_5^* a_6^* \, ,~~
\eea
for the 4-wave ($1 \leftrightarrow  3$) and  the  6-wave ($3 \leftrightarrow  3$) systems respectively.
Here  we use shorthand notations: $a_j\= a _{\B k_j}$ and  $\d _{1,2,3}^{4,5,6}\=\d(\B k_1+\B k_2+\B k_3-\B k_4-\B k_5-\B k_6)$,
${ k_j} = |{\B k_j}|$.

 Statistical description of weakly interacting waves can be reached~\cite{ZLF} in terms of the KE
 \BSE{KE}\BE{KEa}
 \frac{\p n(\B k, t)}{ \p t}= \mbox{St}(\B k, t)\,,
 \ee
 for the waveaction spectrum
 $n(\B k, t)$, defined by
 $
 \< a(\B k, t)a^*(\B k',t)\> = n(\B k, t)\d(\B k-\B k')\,,
$
where $\< \dots \>$ stands for
 the ensemble  averaging.
 The collision integral St$(\B k, t)$ can be found in various ways~\cite{ZLF}, including the Golden Rule widely used in quantum mechanics. For the 4-wave ($1 \leftrightarrow  3$) and  the  6-wave ($3 \leftrightarrow  3$) we have respectively
\BEA{KE13}
\!\! \mbox{St}_{1\!\leftrightarrow 3}& =&\!\!\frac{\pi}{12}\!\int \!\!d\B k_1 \dots d\B k_3 \Big\{  |V_{\,\,\B k}^{\, 1,2,3}|^2  \, \d_{\,\,\B k}^{\, 1,2,3}\,\C N_{\,\,\B k}^{\, 1,2,3}\br
  && \times \d (\o_{\B k}-\o_1-\o_2-\o_3)   \br
 &&   \hskip -1.5 cm +3\, |V_{\,\, 1}^{\,\B k , 2,3}|^2  \, \d_{\,  1}^{\,\B k , 2,3}\,\C N_{\,\, 1}^{\,\B k , 2,3}  \d (\o_1-\o_{\B k}-\o_2-\o_3)\Big \}\,,  \br
 \C N_{\,\, 1}^{\, 2,3,4}&\=&n_1n_2n_3n_4\big( n_1^{-1}-n_{2}^{-1}-n_{3}^{-1}-n_{4}^{-1} \big)\, ;
 \eea
\BEA{KE33}
  \mbox{St}_{3\!\leftrightarrow 3}& =& \frac{\pi}{24}\!\int \!\!d\B k_1 \dots d\B k_5\,|W_{\,\,\B k, 1,3}^{\,4,5,6}|^2  \, \d_{\, \B k, 1,3}^{\, \,4,5,6} \br
  && \hskip -1 cm \times \d (\o_{\B k}+\o_1+\o_2-\o_3-\o_4-\o_5)\,  n_{\B k}n_1n_2n_3n_4n_5\br
 && \hskip -1 cm \times \big( n_1^{-1}+n_{2}^{-1}+n_{3}^{-1}-n_{4}^{-1}-n_{5}^{-1} -n_{6}^{-1}\big)\ .
 \eea
 \ese
Scaling solutions of these KE's (up to a  constant prefactor  $A$),
\BE{sol}
n(k)= A \, k^{-x}\,,
\ee
can be   found under two conditions~\cite{ZLF}:

\textbullet~\emph{Scale-invariance} of the  wave system, when the frequency of waves and the
 interaction coefficients are homogeneous functions of wave vectors:
$\displaystyle
  \o(\l k)=\l^{\a_2} \o(k)\,,\
V(\l \B k_1 ; \l\B k_2 , \l\B k_3 ,\l\B k_4) \=  \l^{\a_4}V(\B k_1 ; \B k_2 , \B k_3 ,\B k_4)\,,
$
 and a similar relationship for $W_{1,2,3}^{4,5,6}$ with an index $\a_6$.

\textbullet~\emph{Interaction locality},   in a sense that the main contribution to the energy balance of a given $k$-wave (with wavevector $k$) originates from its interaction with $k'$-waves with $k'\sim k$. Mathematically it means that all integrals over $k_1$, $k_2$, etc.  in the KE's~\eq{KE} \em converge, \em and therefore
the leading contribution to the collision integral indeed originates from the regions $k_2\sim k$, $k_3\sim k$, etc.
Note that nonlocal spectra are not solutions of the KE's~\eq{KE} and, therefore, physically irrelevant.

  To find the scaling index  $x$ for turbulent spectra with a constant energy flux over scales,
we note that all KE's~\eq{KE} conserve the total energy of the wave system,
$ \displaystyle
 d E\big /dt=0\,, \quad E\=\int E_{\B k} \, d \B k\,, \quad E_{\B k}\= \o_{\B k} \, n_{\B k}$. Therefore the  $\B k$-space
 energy density, $E_{\B k}$,  satisfies a continuity equation:
$\displaystyle \frac{\p E_k}{\p t}+ \frac{\p \ve_k}{\p k}=0\ .
$
Here $\ve_k$ is the energy flux over scales, expressed via an integral over sphere of radius $k$:
 $\displaystyle
\ve_k= \int _{k'<k} d\B k'\,  \o_{k'} \, \mbox{St}(k',t)$.

Under the assumption  of the interaction locality,
 one estimates the $d$-dimensional integral $\int d \B k$ as $k^d$, the interaction coefficients
$V_{\,\,1}^{\,2,3,4}\sim V_{\,\,k}^{\,k,k,k}\sim V k^{\a_4}$,
 $W_{\,\,1,2,3}^{\,4,5,6}\sim W_{\,\,k,k,k}^{\,k,k,k}\sim W k^{\a_6}$ and
$n_k= A_p \, k^{-x_p}$ (for the $p$-wave interactions). Therefore:
 \BSE{sol1}
\BEA{sol1A}
 \label{sol1B}
\ve_k&\sim& k^{3 d} \, ( V  \, k^{ \a_4})^2 \  ( A_4 \,   k^{- x_4})^3  \,, \     1\Leftrightarrow 3\ \mbox{scatering}; ~~~\\ \label{sol1C}
\ve_k&\sim&  k^{5 d} \,  ( W \, k^{ \a_6})^2 \,  ( A_6 \,  k^{- x_6})^5  \,, \    3\Leftrightarrow 3\ \mbox{scattering} .~~~
\eea
For the spectra of turbulence with a constant energy flux $\ve_{k}=\ve=$const., i.e. $\ve_k\propto k^0$. For the  $p$-wave process
this gives
the  scaling exponent of $n(k)$, $x_p$, and   the energy scaling exponent $y_p$, $E(k)\propto k^{-y_p}$:
 \BE{sol1D}
x_p=d+ 2 \a_p/(p-1)\,, \quad y_p=x_p -\a_2\ .
\ee \ese
In fact, these expressions are  valid for any $p>2$. For the 3- and the
4-wave processes (with $p=3$ and $p=4$) this gives the well-known results, see e.g.~\Ref{ZLF}.
Note however, that the 4-wave  $1 \leftrightarrow  3$ is considered here for the first time, and it
 is different from the previously considered  standard $2 \leftrightarrow  2$
processes.

{\bf 3. \emph{6-wave KW turbulence}}.
Finding the effective interaction Hamiltonian ${\C H}_{\hbox{\small int}}$ for KWs appears to be a hard task.
For the 6-wave process, which assumes that the underlying vortex is perfectly straight, this task was
accomplished only recently \Refs{LNR-1}.
Effective $3\!\leftrightarrow 3$-interaction coefficient $W$ was shown to have a form
\BE{W6}  W_{\,1,2,3}^{\,4,5,6} =  - {3\, \B k_1\B k_2\B k_3\B k_4\B k_5\B k_6}\,  F_{\,1,2,3}^{\,4,5,6}\big / {4\pi \kappa} \, \,,
\ee
where $F $ is a non-singular dimensionless function of $\B k_1\,, \dots \B k_6$,  close to
  unity in the relevant region of its arguments (KW case is 1D, but we still use boldface for the wavevectors, reserving non-bold
  notation as $k_j=|\B k_j|$).
Notice that the form of  \Eq{W6} could be expected because  it
   demonstrates a very simple physical fact: long KWs (with small $k$'s) can contribute to the energy of a  vortex line only when they produce curvature. The curvature, in turn,  is proportional to wave amplitude $a_k$ and, at fixed amplitude, is inversely proportional to their wave-length, i.e. $\propto k$.  Therefore in the effective motion equation each $a_j$ has to be accompanied by $k_j$, if $k_j\ll k$. Exactly this statement is reflected by \Eq{W6}.
   One can say that cumbersome calculations~\cite{LLNR-09} support these reasoning, and additionally provide  with
an explicit expression for $F$.

Equation~\eq{W6}   estimates $ W_{\,1,2,3}^{\,4,5,6}$ as $W k^6$. Thus, \Eq{sol1C}  reproduces the Kozik-Svistunov (KS) scaling
for the ${3\!\leftrightarrow 3}$ processes, which for further discussion is   writhen with a dimensionless constant $C\Sb{KS}$:
 \Fbox{\vskip -.4 cm \BEA{KS}
 && n\Sb{KS} =
  \frac{ C\Sb{KS}   \k^{2/5} \ve^{1/5}}  {  k^{17/5}} \
   \Rightarrow \
  E\Sb{KS} = \frac{ C\Sb{KS} \L   \k^{7/5} \ve^{1/5}}  {  k^{7/5}}\,,  \\ \nn
  && \mbox{ Nonlocal $(3  \!\leftrightarrow \!3) $ Kozik-Svistunov (KS) spectrum.}\eea
  \vskip -.4cm  }

{\bf 4.~\emph{Nonlocality of  the 6-wave KW theory}}.
To test locality of the KS spectrum \eq{KS},
let us consider the  $3 \!\leftrightarrow \!3$ collision term~\eq{KE33} for KW with the interaction amplitude  $  \C W_{\,1,2,3}^{\,4,5,6}$ as in~\eq{W6} and $n(k)$ as in \Eq{sol}.
 In the    IR region  $k_1\ll k, k_j$, $j=2,3,4,5$, we have $F \to 1$ and
 the integral over $\B k_1$ scales as:
  \BE{int}
\Psi\= \frac 2\kappa\int _{1/\ell}    k_1^2\,  n(k_1\!)\,  d k_1 = \frac {2\, A}{\kappa} \int _{1/\ell}   k_1^{2-x} d k_1 \ .
\ee
   Lower limit $0$ in \Eq{int} is replaced by $1/\ell$, where $\ell$ is  the  mean inter-vortex separation $\ell$, at which approximation of non-interacting vortex lines fails and one expects a cutoff of the power like behavior~\eq{sol}.
 Prefactor 2 in \Eq{int} reflects the fact that the ranges of positive and  negative $\B  k_1$ give equal contributions, and factor
  $1/\k$  is introduced to make parameter $\Psi$ dimensionless.  $\Psi$ has a meaning of the mean-square angle of the deviation of the vortex lines from straight.   Therefore $\Psi \lesssim 1$; for highly polarized vortex lines $\Psi \ll  1$.

Clearly, integral~\eq{int} IR-diverges if $x>3$, which is the case for the KS spectrum~\eq{KS}  with $x_6=17/5$.    Note that all the similar integrals over $\B k_2$, $\B k_3$, $\B k_4$, and $\B k_5$ in \Eq{KE33} also diverge exactly in the same manner as integral~\eq{int}.
Moreover, when two of the wavenumbers belonging to the same side in the sextet tend to zero simultaneously then
each of such wavenumbers will yield an integral as in \eqref{int}, and the net result will be the product
of these integrals, i.e. a stronger singularity than in the case of just one small wavenumber.
On the other hand, small wavenumbers which are on the opposite sides of the resonant sextet do not lead to
a stronger divergence because of an extra smallness arising in this case in \eqref{KE33} from
$( n_1^{-1}+n_{2}^{-1}+n_{3}^{-1}-n_{4}^{-1}-n_{5}^{-1} -n_{6}^{-1})$.

Divergence of the integrals in \Eq{KE33} means that KS-spectrum~\eq{KS}
   is not a solution of the
   KE~\eq{KE33} and thus non-realizable. One should find another, self-consistent solution of this KE.

{\bf 5.~\emph{Effective  4-wave theory}}.
Thus, the strongest nonlocality of the 6-wave theory arises from those sextets that
  contain, on the same sextet side, two small wavenumbers with $k_j \lesssim 1/\ell$.
Thus the 6-wave resonance conditions
$\o_k+\o_1+\o_2=\o_3+\o_4+\o_5\,, \ \B k+\B k_1+\B k_2=\B k_3+\B k_4+\B k_5\,, $
 effectively become
 \bea \nn \B  k=\B k_1+\B k_2+\B k_3, && \B k_1=\B k+\B k_2+\B k_3, \ \\ \label{5}
\B  k_2=\B k+\B k_1+\B k_3, &&\mbox{or} \;\;\; \B  k_3=\B k+\B k_2+\B k_1,
\eea
and respective conditions for the frequencies, which implies a 4-wave process of the ($1\leftrightarrow 3$)-type.
In the other words, one can interpret such nonlocal sextets on straight vortex lines as quartets on curved vortices, with
the slowest modes in the sextet responsible for the large-scale curvature $R$ of the underlying vortex line in the 4-wave
approach.

 To derive an effective 4-wave KE, let us start with the 6-wave collision integral \eqref{KE33} and  find the
 leading contributions to it when the spectrum $n_k$  is steeper than $k^{-3}$ in the IR-region.
  There are four of them. The first one originates from the region where $k_1$ and $k_2$ are much smaller than the rest of $k_j$'s.  The three other contributions originate from the other side of the sextet: regions where either $k_3$ and $k_4$, or $k_3$ and $k_5$, or  $k_4$ and  $k_5$ are small. These contributions are equal and we may find only one of them and multiply the result by three.
 Notably,   the  sum of the four  contributions can be written  exactly in the form of the $(1\!\leftrightarrow\!3)$-collision term~\eq{KE13} with the effective $(1\!\leftrightarrow\!3)$-interaction amplitude
\BE{W4}
 V_{\,\,1}^{\, 2,3,4}=-3\Psi \, \B k_1\B k_2\B k_3\B k_4\big /  (4\pi \sqrt{2}) \,,
\ee
because, as shown in~\cite{LLNR-09}, $\displaystyle \lim _{k_1\to 0}F(\B k_1,\B k_2,\B k_3|\B k_4\B k_5,\B k_6)= 1$.  Deriving \Eqs{KE13} with $ V_{\,\,1}^{\, 2,3,4}$, \Eq{W4},  we took only leading contributions in the respective IR regions, factorized the integrals over these wave vectors like in~\Eq{int} and
 took only the zeroth order terms with respect to the small wavevectors (by putting these wavenumbers to zero) in the rest of the expression~\eq{KE33}.

Equation  \eqref{KE13} with $V_{\,\,1}^{\, 2,3,4}$ as in  \Eq{W4} is an effective 4-wave KE, which we were aiming to obtain. This KE corresponds to   interacting quartets of KWs  propagating along a
vortex line having a random large-scale curvature $R \lesssim \ell$.
 Equation~\eq{W4}  estimates $V_{\,1}^{\,2,3,4}$ as $\C V  k^4$ with $\C V\sim \Psi $. Using this scaling in \Eq{sol1B}, we
 arrive at a spectrum for the ${1\!\leftrightarrow 3}$ processes with  scaling exponents $x_4=11/3$ and $y_5=5/3 $,
 \Fbox{\vskip -.4 cm \bea
&& n\Sb{LN}= \frac{C\Sb {LN}\, \ve^{1/3}}{ {\Psi^{2/3}} k^{11/3}}  \Rightarrow
 E\Sb{L N}   =  \frac{C\Sb {LN} \L \, \k \, \ve^{1/3}}{ {\Psi^{2/3}}  k^{5/3}},
 \label{LN}
 \\  && \hskip - .3 cm \mbox{Local $(1  \!\leftrightarrow \!3) $ L'vov-Nazarenko (LN) spectrum.} \nn
  \eea \vskip -.4 cm
  }

\noindent
{\bf 6.~\emph{Locality of the 4-wave LN spectrum}}.

 Clearly, for the new spectrum  \eqref{LN} to be a valid solution, it must satisfy the locality test.
 Thus let us substitute this spectrum into the ${1\!\leftrightarrow 3}$ collision integral
 \eqref{KE13} and check it convergence. Leaving details of the locality test to the
 online supplement to this Letter, we just outline here the main steps.
 Taken separately, the first and the second terms in the curly bracket yield IR divergent
 integrals, but when taken together the leading order singularities of these terms
 cancel with one another, and the net result is a IR convergent integral.
 In the ultraviolet  (UV)  region, the singularity should only be checked in the second term of the curly bracket of~\eq{KE13},
because the first term of this bracket does not have a UV region due to the $\omega$ $\delta$-function.
Also because of this $\delta$-function, in the UV range of the second term two of the $\B k$'s, say $\B k_1$ and $\B k_2$, must be large
simultaneously so that  $\B k_1 \simeq \B k_2$, which leads to a UV convergent integral.
Thus, the LN spectrum \eqref{LN} appears to be local and, therefore, it is a valid solution for describing the KW turbulence.



\noindent
{\bf  7. \emph{Conclusions}}.  \\ \noindent
 \textbullet~We presented a new effective 4-wave theory of KW  turbulence consisting of wave quartets interacting on vortex lines with random large-scale curvature. We  derived   an effective 4-wave KE,
 \eqref{KE13}, \eqref{W4}, and solved it to obtain a new KW spectrum \eqref{LN}. We proved that this   spectrum is local, and therefore it is a valid solution of the KE, which should replace the nonlocal (and therefore invalid)  KS-spectrum \eqref{KS} in the theory
 of quantum turbulence. In particular, it is now necessary to revise the theory of the classical-quantum crossover scales and its predictions for the turbulence dissipation rate~\cite{LNR-1,KS-04}.
   Further, a similar revision is needed for the analysis of laboratory experiments and numerical simulations of superfluid turbulence, which have been done over the last five years with reliance on   the un-physical KS spectrum~\eq{KS}.

\textbullet~The   difference between  the LN-exponent $-5/3$ (see ~\eq{LN}) from the KS-exponent $-7/5$ (see~\eq{KS}) is  4/15 which is rather small. This  may explain why the previous numerical experiments seem to agree with
the KS spectrum, obtained numerically in \cite{DNS}. However, by inspection one can also see that these results also agree with the LN slope.
Differences in physical processes corresponding to the KS and LN spectra, result in different   dimensional prefactors in these spectra, in particular the different dependence on the energy flux $\varepsilon$, as well as an extra dependence on the large-scale
behavior (through $\Psi$) in \eq{LN}. Careful examination of such prefactors is necessary in future numerical simulations in order to
 resolve uncertainties related to close spectrum exponents and thereby test the predicted dependencies. Such numerical
 simulations can be done efficiently with the Local Nonlinear Equation (LNE) suggested in \cite{LLNR-09} based on the detailed
 analysis of the nonlinear KW interactions:
 \Fbox{
 \vskip -.4 cm
 \bea\label{LNE}  i \, \frac{\p  \widetilde{w}}{\p t}+ \frac{\kappa}{4\pi}\, \frac{\p}{\p z}\left[ \left(\L - \frac 14 \, \left | \frac{\p \widetilde{w} }{\p z }\right |^4 \right) \frac{\p  \widetilde{w} }{\p z } \right]=0\,. ~~~~ \\ \nn
\mbox{LN equation for KWs.}~~~  \eea
\vskip -.4 cm}
The LNE model is similar but not identical to the Truncated LIA model of \cite{DNS} (these models become asymptotically identical
for weak KWs).

\textbullet~ The KS and LN prefactors   contain
very different numerical constants $C$: an order-one constant in LN ($C \Sb{LN}\sim 1$, yet to be found)
and a zero constant in KS  ($C\Sb {KS} = 0$ as  a formal consequence of its nonlocality). Also we should note a
mysterious very small numerical factor $10^{-5}$  in formula (12) for the energy flux in
Ref.~\cite{KS-04}, that has no physical justification.  Actually, nonlocality of the energy transfer over scales means that this number should be very large, rather than
very small.  This emphasizes the confusion, and highlights the need for numerical
re-evaluation of the spectrum's prefactor.

\textbullet~ Obviously, the differences between
 the KS and the LN spectra, in the exponents and, most importantly, in the prefactors,
 is important for practical  analysis and interpretations of experimental data.
  At the same time, the difference  between the underlying physics of  the local and the nonlocal energy cascades,
 is  important from the fundamental, theoretical viewpoint.

\textbullet~In this work, the effective local 4-wave KE was derived from the 6-wave KE by exploiting nonlocality of the latter which is valid only when the 6-wave KE is valid, i.e. when all the scales are
weakly nonlinear, including the ones at the IR  cutoff. However, The resulting 4-wave KE is likely to be applicable more widely, when only the small scales, and not the large scales, are weak. A similar picture was previously observed for the nonlocal turbulence of Rossby/drift waves in \cite{nazPHD} and for nonlocal MHD turbulence in \cite{nonlocMHD}. In future we plan to attempt derivation of the 4-wave KE directly from the dynamical equations for the KWs, which would allow us to extend its applicability to the case with strong large scales.

\textbullet~Finally we note that the suggested here theory can potentially be useful for other one-dimensional  physical systems,  including  optical fibers, where nonlinear interactions of one-dimensional wave packages becomes important with increase in network capacity.

\vskip .1cm
\noindent
{\bf  \emph{Acknowledgements}}:
We thank J.~Laurie and O.~Rudenko for help in the evaluation of the effective interaction coefficient.
We acknowledge  support of  the US-Israel Binational Scientific Foundation administrated by the Israeli Academy of Science, and of the
EC --
Research Infrastructures under the FP7 Capacities Specific Programme,
MICROKELVIN project number 228464.

\newpage
{\bf  \emph{Appendix: Proof of locality of the  energy transfer in the $\bf (1 \! \leftrightarrow \! 3) $-wave processes}}. \vskip 0.4cm

Mathematically, locality of the  energy transfer in the $\bf (1 \! \leftrightarrow \! 3) $-wave processes means convergence of the multi-dimensional integral in the corresponding collision term~\eq{KE13}. Here we will show that proof of convergence in \Eq{KE13} is a delicate issue and cannot be done only on the basis of power counting because the latter would give a  divergent answer. \vskip 0.2cm

\paragraph{Proof of the infrared (IR) convergence. } Let us show that in the IR region, when at least one of the wave vectors, say $k_2$,  is much smaller then  $k$,  only
a quadruple cancelation of the largest, next to the largest and the two further sub-leading  contributions  appear to result in the final, convergent  result for the collision term~\eq{KE13}.

Three integrations in  \Eq{KE13} are restricted by two conservation laws, namely by
\bse\label{ap1}
\bea\label{ap1A}
1\rightarrow3:&& \B k=\B k_1+\B k_2+\B k_3\,, \  k^2= k_1^2+ k_2^2+ k_3^2\ ;~~~~~
\eea
in the first term and by
\bea\label{ap1B}
3\rightarrow1:&& \B k+\B k_2+\B k_3=\B k_1\,, \  k^2 +k_2^2+k_3^2=k_1^2\ ; ~~~~~
\eea \ese
in the second term.
Therefore,
only one integration, say with respect to $\B k_2$,  remains  in each term.

In the IR
region $k_2\ll k_1\ll k$,
 we find from \Eq{ap1}  for  the $(1 \rightarrow 3)$ and the $(3 \rightarrow 1)$ terms:
 \vskip - .3cm
\bse\label{ap2}
\bea\label{ap2A}
1\rightarrow3:&& \B k_1=\B k  - \frac{k_2^2}{\B k_1+\B k_2} \approx \B  k -  \frac{k_2^2}{\B k} \,,  \\ \nn
1\rightarrow3:&&\B  k_3=- \frac{\B k_1\B  k_2}{\B k_1+\B k_2} \approx - \B k_2  \,,  \\
 \label{ap2B}
3\rightarrow 1:&& \B  k_1=\B k + \frac{k_2^2}{\B k+\B k_2} \approx \B k +  \frac{k_2^2}{\B k}\,,  \\ \nn
3\rightarrow 1:  && \B k_3= -  \frac{\B k\, \B k_2}{\B k+\B k_2} ~~~\approx -\B k_2 \ .
\eea
\ese
 These equations demonstrate  three important facts:
 \begin{enumerate}
 \item in both cases in the leading order $k_3\simeq  k_2$, i.e. when $k_2\ll k$ then $k_3$ is small as well;~
 \item  the difference between $\B k_1$ and $\B k$ is of the second order in small $k_2$: $|\B k_1-\B k|\simeq k_2^2 /k\,;$~
 \item  these leading contributions to    $ (\B k_1-\B k)$ have the same modulus and different sign in  the
 $(1 \leftrightarrow 3)$-term and in the $(3 \leftrightarrow 1)$-term.
\end{enumerate}
 Therefore in the leading order the expressions for $\C N$ in \Eq{KE13} can be written as:
 \bse \label{rel}\begin{eqnarray}
 \C N_{k}^{1,2,3} \simeq   - x  ( {k_2}/k   )^2 n_k n_2 n_3
 &\simeq &  -    \frac{x\, A^3 }{k^ {\, (x+2)} } k_2^{2(1-x)} \,,~~~~~~  \\
 \label{ap2B}
  \C N_{1}^{k,2,3} \simeq   + x  ( {k_2}/ k   )^2 n_k n_2 n_3
 &\simeq &  +    \frac{x\, A^3 }{k^ {\, (x+2)} } k_2^{2(1-x)} \,,~~~~~~
\end{eqnarray}\ese
where we substituted $n_j$ from \Eq{sol}. Importantly, these estimates
  (in the leading order)   have \emph{the same magnitude and different signs}.

 Next step is to compute  integrals
  \bse\label{ap4}
\bea\label{ap4A}
 &&  I_{1\rightarrow3} \= \int d\B k_1 d\B k_3 \d (\B k-\B k_1-\B k_2-\B k_3)\\ \nn
 && \times \d ( k^2 -k_1^2 -k_2^2-k_3^2)= \frac{|\B k+\B k_2 |}{2|k^2+2\, \B k \B k_2 - k_2^2|}\rightarrow \frac 1 {2k}\,; \\
 \label{ap4B}
  && I_{3\rightarrow1}  \= \int d\B k_1 d\B k_3 \d (\B k+\B k_2+\B k_3-\B k_1)\\ \nn
 && \times \d ( k^2 +k_2^2+k_3^2-k_1^2)=  \frac 1 {2|\B k+\B k_1 |}  \rightarrow \frac1  {2k}\,,
 \label{Ap2B}   \eea
\ese
  i.e. in the leading order these results coincide and do not contain the smallness.

Now we can find the contributions to St$_{1\leftrightarrow 3}$, given by \Eq{KE13}, from the region $k_2\ll k$. According to \Eq{W4}   we can write  $V_k^{1,2,3}=V_1^{k,2,3}= V \B k\, \B k_1\B k_2\B k_3$. Using our estimates~\eq{rel}  for $\C N $ and \Eq{ap4} we have:
 \bse\label{ap5}
\bea\label{ap5A}
1\rightarrow3:\  \mbox{St}_{1\rightarrow3}^{k_2\ll  k}&\approx& - \frac{x \pi V^2\, A^3}{24 k^{x-1}}\int k_2^{2(3-x)}d\B k_2 \, ;     \\
 \label{ap5B}
3\rightarrow1:\    \mbox{St}_{1\rightarrow3}^{k_2\ll  k}&\approx& +  \frac{3\, x \pi V^2\, A^3}{24 k^{x-1}}\int k_2^{2(3-x)}d\B k_2 \ . ~~~~   \eea
\ese
One can see that, in spite of the deep cancelations in the estimates for $\C N $, the integrals~\eq{ap5} diverge if $x\ge 3.5$, which is
satisfied for LN-scaling exponent  $x=11/3$.

Nevertheless on has to take into account the following:  the $(1\rightarrow3)$-contribution to the collision integral has three identical divergent regions: $k_2 \sim  k_3 \ll k_1\approx  k$, $k_1 \sim  k_3 \ll k_2\approx  k$ and $k_2 \sim k_1 \ll k_3\approx  k$, and \Eq{ap5A} estimates only the first one. Therefore the total contribution is
\bse\label{Ap5}
\be\label{Ap5A}
 \mbox{St}_{1\rightarrow3}\Sp{IR} =3 \,  \mbox{St}_{1\rightarrow3}^{k_2\ll  k} \approx  - \frac{3\, x \pi V^2\, A^3}{24 k^{x-1}}\int k_2^{2(3-x)}d\B k_2 \,,
 \ee
 while the $(3\rightarrow1)$-contribution has   only one divergent region $\B k_1\approx \B k$. Therefore,
\be  \label{Ap5B}
 \mbox{St}_{1\rightarrow3}\Sp{IR}=    \mbox{St}_{1\rightarrow3}^{k_2\ll  k} \approx  +  \frac{3\, x \pi V^2\, A^3}{24 k^{x-1}}\int k_2^{2(3-x)}d\B k_2 \,, ~~~~   \ee
i.e. exactly the same result as in \Eq{ap5A}, but with the different sign. Therefore the divergent contributions~\eq{ap5} cancel each other and one has to take into account the next order.

Notice that next order terms in the expansion over $k_2\ll k$ results in the already \emph{convergent } integral
  \be  \mbox{St}_{1 \leftrightarrow3}\Sp{IR}  \propto  \int\limits_0^{k_{\,\rm IR}  \ll  k}  \B k_2 k_2^{(6-2x)} d\B k_2  \,,\ee
 with the LN exponent $x=11/3$.
 \ese
 Moreover, typically excitation of  KWs is symmetrical in $\B k \leftrightarrow -\B k$. In this case, this integral   has an odd integrand and, therefore, it is
 equal to zero.  Then the leading contribution  to  the $(1 \leftrightarrow 3)$-collision term in the IR region   can be summarized as   follows:
  \Fbox{\vskip -0.4cm
\bea  \nn
&&\hskip - .5 cm   \mbox{St}_{1 \leftrightarrow3}\Sp{IR} \sim  \frac{ V^2\, A^3}{ k^{x+1}} \int\limits_0^{k_{\,\rm IR}  \ll  k}   k_2^{2(4-x)}d\B k_2 \propto    k \Sb{\, IR}^{\, 9-2x }  \ .      \\  \label{ap6}
 && \mbox {The IR convergence require:}\quad   x<\frac 92\ .
  \eea
 \vskip -0.4cm}
 With LN exponent $\displaystyle x=\frac 9 3$ this gives St$_{1 \leftrightarrow3}\Sp{IR} \propto  k\Sb {IR}^{5/3}\=  k\Sb {IR}^{\delta_{\rm IR}}$. Here we introduce
  an  ``IR convergence reserve": $ \displaystyle  \delta\Sb {IR}= \frac 53$.
\vskip 0.4cm

\paragraph{Proof of the ultraviolet (UV)  convergence. }
 Convergence of the integral~\eq{KE13} in the  UV region, when one of the wave vectors, say
 $k_2\gg k$,
  can be established i a similar manner.

  Notice first of all that in the $1 \rightarrow 3$ term in  \Eq{KE13},  there is no UV region, because by the 2$\,\sp{nd}$ of \Eq{ap1A} we have
    $k_j\le k$.
    In the $3 \rightarrow 1$ term to satisfy \Eq{ap1B}  in the leading order   we can take $\B k_{2}\simeq \B k_1; \;\; k_2  \geq k\Sb {\, UV}\gg k$ (case $\B k_{3}\simeq \B k_1; \;\; k_3 \geq k\Sb{\, UV}$ gives
    an identical result).  Using parametrization $\B k_1=\B k + {k_2^2}/(\B k+\B k_2) ,  \;\;\B  k_3= -  {\B k\, \B k_2}/(\B k+\B k_2)$ (cf. \eqref{ap2B}) we
get some cancelations in $\C N _1^{k,2,3}$ and the leading order result is
\be
\C N _1^{k,2,3}\propto x(x+1) \left( \frac{k_2}{k}\right)^{-2-x} -  \left( \frac{k_2}{k}\right)^{-2x}\ .
\ee
 Further, similarly to \Eqs{ap4}, one gets $ I_{3\rightarrow 1}\simeq 1/k_2$.
 As before, the interaction coefficient $V\propto k_2^2$ or  $V^2\propto k_2^4$. Counting the powers of $k_2$ one gets:
 \Fbox{\vskip -0.4cm
 \bea \nn
 && \mbox{St}_{1 \leftrightarrow3}\Sp{UV}
  \propto k\Sb{\, UV}^{y }\,, \ y=\max(-2x+4,-x+2) \ . \\  \label{ap6}
  &&\hskip -.5cm \mbox {The UV convergence require } \
   y < 0 \  \Rightarrow x>2\ .  ~~~~~~
    \eea
 \vskip -0.4cm}
 One concludes that in the  case $\displaystyle x= \frac{11}3$,\ St$_{1 \leftrightarrow3}\Sp{UV} \propto k\Sb {UV}^{- 5/3}\= k\Sb {UV}^{-\delta _{\rm UV}},$ where we introduce an ``UV convergence reserve" $\displaystyle \delta\Sb{UV}=\frac 53\ .$

 Notably, $\delta\Sb {IR}= \delta \Sb {UV}$. This equality is not occasional. Observed  ``counterbalanced" IR-UV locality is a consequence of the scale-invariance of the problem. Indeed, for a given values of $k\Sb{\, IR}\ll \~ k \ll k\Sb{\, UV}$ the  IR-energy flux $ k\Sb {IR} \Rightarrow \~k$ (\emph{from}  the IR region  $k \leq k\Sb{\, IR}$ \emph{toward} the region $\sim \~k$)   should scale with  $(k\Sb{\, IR}/\~k) $ exactly in the same manner as the UV-energy flux $\~k \Rightarrow k\Sb {\, UV}$ (\emph{from} the $\~k$-region \emph{toward} the UV-region $k\le k\Sb{UV}$) scales with $ \~k/k\Sb{\,UV}$. This is  because the UV-flux $\~k \Rightarrow k\Sb {\, UV}$ \emph{from } $\~k$-region can be considered   as the IR flux \emph{toward} $k\Sb{\, UV}$-region. Remembering that the IR-energy flux $ k\Sb {IR} \Rightarrow \~k$   scales like 
  $\displaystyle \big( k\Sb{\,IR}/\~ k\big )^{\delta_{\rm IR}}$, while the UV-flux $\~k \Rightarrow k\Sb {\, UV}$ is proportional to 
    $\displaystyle \big (\~k/k\Sb{\, UV}\big )^{\delta_{\rm UV}} $, one immediately concludes that  $\delta\Sb {IR}$ should be equal to $\delta \Sb {UV}$.

 \vskip .2 cm

The overall conclusion is that the collision term St$_{1 \leftrightarrow3}$ is convergent in both the IR and the UV regions for  $\displaystyle x=\frac {11}3$
and the \emph{energy transfer in the   $1 \leftrightarrow3$ kinetic equation is local}.

\end{document}